# Role of spin-phonon and electron-phonon interactions in phonon renormalization of $(Eu_{(1-x)}Bi_x)_2Ir_2O_7$ across the metal-insulator phase transition: Temperature-dependent Raman and X-ray studies

Anoop Thomas [1], Prachi Telang [2], Kshiti Mishra [2], Martin Cesnek [3], Jozef Bednarcik [4 #], D V S Muthu [1], Surjeet Singh [2] and A K Sood[1*]

[1]*Department of Physics, Indian Institute of Science, Bangalore, 560012, India*

[2] *Department of Physics, Indian Institute of Science Education and Research, Dr Homi Bhabha Road, Pune 411008, India*

[3]*Department of Nuclear Reactors, Faculty of Nuclear Sciences and Physical Engineering, Czech Technical University in Prague, V Holesovickach 2, 180 00 Prague, Czech Republic*

[4]*Deutsches Elektronen-Synchrotron DESY, Notkestrasse 85, D-22607 Hamburg, Germany*

## Abstract

We report temperature-dependent Raman scattering and X-ray diffraction studies of pyrochlore iridates, $(Eu_{(1-x)}Bi_x)_2Ir_2O_7$, for x=0, 0.02, 0.035, 0.05 and 0.1. The temperature variation in Raman experiments spans from 4 K to 300 K, covering the metal-insulator phase transition accompanied by paramagnetic to all-in/all-out (AIAO) spin ordering ($T_N$). These systems also show a Weyl semi-metal (WSM) phase at low temperatures (below ~50 K). We show that the Ir-O-Ir bond bending mode, $A_{1g}$ (510 cm$^{-1}$), shows anomalous softening in the magnetically ordered AIAO state, arising primarily from the spin-phonon interaction due to the phonon-modulation of the Dzyaloshinskii-Moriya (DM) spin-exchange interaction. The two stretching modes, $T_{2g}^1$ (307 $cm^{-1}$) and $T_{2g}^2$ (382 $cm^{-1}$) harden significantly in the magnetic insulating phase. The $T_{2g}$ phonons also show anomalous temperature dependence of their mode frequencies, hitherto unreported, due to strong electron-phonon coupling. The signatures of the WSM state are observed in phonon renormalization below 50 K due to strong electron-phonon interaction. Our experimental results establish strong magneto-elastic coupling below $T_N$ and significant electron-phonon interactions in the metallic phase above $T_N$ as well as in the low-temperature WSM state.

**KEYWORDS**: pyrochlore, spin-phonon coupling, electron-phonon interaction, metal-insulator transition, Dzyaloshinskii-Moriya interaction, Weyl semi-metal, phonon renormalization.

*****Corresponding Author**,** E-mail: asood@iisc.ac.in

[#]Current affiliation - Institute of Physics, Faculty of Science, P.J. Safarik University in Kosice, Park Angelinum 9, 041 54 Kosice, Slovak Republic

## 1 INTRODUCTION

There has been an upsurge of interest in understanding exotic properties of pyrochlore iridium oxides ($A_2Ir_2O_7$, A= lanthanide, Bi, Pb, or Y) due to large spin-orbit coupling (SOC) arising from heavy iridium atoms, intermediate electron correlations and geometry-induced spin frustration [1–4]. Heisenberg antiferromagnetic isotropic exchange (IE) interactions make the iridium ($Ir^{4+}$) effective spin 1/2 on a tetrahedral motif frustrated, preventing the spins to have a long-range magnetic ordering. However, the Dzyaloshinskii-Moriya (DM) interactions arising from strong SOC competes with Heisenberg interactions to result in an ordered antiferromagnetic structure with all-in/all-out (AIAO) ground state below the transition temperature $T_N$ in pyrochlores with small ionic radii A such as Lu, Yb, Ho, Y, Dy, Tb, Gd, Eu, Sm and Nd. At low temperatures, DM interaction wins over Heisenberg interaction in the iridates having smaller $A^{3+}$ size resulting in a magnetic insulating phase. The electron hopping arising from the chemical pressure of the $A^{3+}$ site can suppress this magnetic ordering in $Pr_2Ir_2O_7$ and $Bi_2Ir_2O_7$ at low temperatures [5–7]. On the other hand, the paramagnetic to antiferromagnetic phase transition in three iridates, $Eu_2Ir_2O_7$, $Sm_2Ir_2O_7$, and $Nd_2Ir_2O_7$, is accompanied by a metal-insulator transition (MIT) without change of crystal symmetry [8–10]. Theoretical calculations show that below the MIT temperature, topological phases like Weyl semi-metal (WSM), axion insulator and topological Mott insulator are possible [2,3,11]. The experimental probes to show the topological phases in pyrochlore iridates include optical conductivity, resistivity, and angle-resolved photoemission spectroscopy (ARPES) along with varying pressure, temperature and doping [12–16]. The WSM state at low temperatures has been seen in optical conductivity measurements in $Eu_2Ir_2O_7$ by way of the linear temperature dependence of optical conductivity below Neel temperature $T_N$ and inter-band optical conductivity vanishing at low frequencies [12]. The Drude spectral weight was found to be independent of temperature in the metallic phase, whereas it decreased smoothly in the AIAO phase. Simultaneous pressure and temperature-dependent electrical transport experiments have also shown a quantum critical point in $Eu_2Ir_2O_7$below 120 K and above 6 GPa [13].

Similar to physical pressure, quantum criticality is observed with chemical doping also. Doping can be done by substitution of $A^{3+}$ site or $Ir^{4+}$site atoms. The substitution of $Ir^{4+}$ with $Rh^{4+}$and $Ru^{4+}$ reduces SOC strength [17–19] whereas isovalent substitution in $A^{3+}$ atoms results in tuning of the electronic bandwidth. Since the topological states arise from $Ir^{4+}$ electrons, the study of electronic and magnetic phases is richer when the non-magnetic $A^{3+}$ ion is substituted. Telang *etal.* [16] have studied emerging electronic and magnetic phases in $Eu_2Ir_2O_7$ with bismuth (Bi) doping. Since the lattice constant of $Bi_2Ir_2O_7$ (10.3307 Å) is higher than $Eu_2Ir_2O_7$ (10.2989 Å), the isovalent substitution of $Eu^{3+}$ with $Bi^{3+}$ should increase the lattice parameter in $(Eu_{(1-x)}Bi_x)_2Ir_2O_7$ with x as per Vegard's law. In contrast, an anomalous lattice volume decrease is observed in $(Eu_{(1-x)}Bi_x)_2Ir_2O_7$ for x=0.02 to 0.035. Further, resistivity for these systems at low temperatures follows $1/T^{\alpha}$, suggesting the phase to be close to the WSM phase [16]. For x=0.02, $\alpha$ ~1 and combining the transport data with the measurements of the Seebeck coefficient, it has been proposed that the low-temperature phase is WSM, with the Fermi energy close to the Weyl nodes. For x=0, Fermi energy is below the Weyl nodes (i.e. hole-doped), whereas for x=0.035, Fermi energy is above the Weyl nodes (i.e. electron-doped) [16].

In recent years, the role of electron-phonon interaction has been recognized to be important in understanding electronic properties and phase behaviour of pyrochlore iridates [20]. Another interaction found to play a crucial role in transition metal oxides is the spin-phonon coupling which has led to the observation of phenomena, such as the thermal Hall effect [21,22], stabilization of magnetic order with epitaxial strain [23,24] and spintronics [25]. Raman [18,26] and infrared [27] studies on some of the pyrochlore iridates show phonon anomalies at and below $T_N$. For example, Raman studies on $Eu_2Ir_2O_7$ [26] show an anomalous softening and lineshape anomalies of the Ir-O-Ir bending vibration, $E_g$ mode, below $T_N$, attributed to the strong coupling of the $E_g$ phonon to spin, charge, and orbital excitations. Notably, they did not see any anomaly in the other Ir-O-Ir bending vibration of the $A_{1g}$ symmetry or in the remaining Raman allowed modes. Recent infrared studies of $Y_2Ir_2O_7$ [27] have brought out that the phonon softening of Ir-O-Ir bending related vibrations arises due to spin-phonon coupling mediated by the DM interaction rather than the commonly considered isotropic exchange interaction. Motivated by these experiments, we have done a detailed temperature-dependent Raman study of $(Eu_{(1-x)}Bi_x)_2Ir_2O_7$ (x=0, 0.02, 0.035,0.05 and 0.1) and show that the $A_{1g}$ mode associated with Ir-O-Ir bending vibration and the other two low-frequency $T_{2g}$ modes show phonon anomalies which have not been reported so far in literature. Further, the $T_{2g}$ modes show anomalous hardening (as temperature increases) above $T_N$ due to electron-phonon interaction. Notably, signatures of the WSM phase are seen below 50 K in the temperature dependence of phonon frequencies and linewidths.

## 2 METHODS

Polycrystals of $(Eu_{(1-x)}Bi_x)_2Ir_2O_7$ were synthesized by solid-state reaction route and characterized by high-resolution X-ray synchrotron diffraction, electrical resistivity, thermoelectric power, muon spin resonance and specific heat as reported in our recent work [16]. The MIT, accompanied by magnetic order, is observed in samples with x=0, 0.02, 0.035 and 0.05 at temperatures 122, 115, 108 and 63 K, respectively. For x=0.1, temperature-driven MIT is suppressed, showing metallic behaviour throughout the entire temperature range. The low-temperature X-ray diffraction experiments between room temperature and 90 K were carried out on series of composition $(Eu_{(1-x)}Bi_x)_2Ir_2O_7$ (x=0, 0.02, 0.05 and 0.1) using high-energy photons (λ=0.020713 nm) at the Beamline P02.1 of Petra III, DESY in transmission mode [28]. Low-temperature micro-Raman studies (4 K to 300 K) were carried on sintered pellets of $(Eu_{(1-x)}Bi_x)_2Ir_2O_7$ (x=0, 0.02, 0.035, 0.05 and 0.1) using continuous flow helium cryostat (M/s Oxford Instruments, Model Microstat) with temperature controller ITC502. After attaining the set temperature, we waited for 10 min before recording the spectra. Raman spectra were recorded using Horiba LabRam spectrometer in back-scattering geometry, using 50X objective and laser excitation of 532 nm with laser power of ~0.4 mW on the sample. Raman spectra were fitted with sum of Lorentzian functions to extract frequency and linewidth (full width at half maximum (FWHM)) of individual Raman modes.

# 3 RESULTS

## 3.1 X-ray diffraction as a function of temperature and doping

Figures S1(a) and (b) show typical X-ray diffraction patterns along with the Rietveld fitting at 300 K and 90 K for x=0 and x=0.02. As no new Bragg peak is seen at low temperatures for all values of x, this confirms the absence of structural change with doping and with temperature, particularly below $T_N$. The temperature dependence of lattice parameters of $(Eu_{(1-x)}Bi_x)_2Ir_2O_7$ (x=0, 0.02, 0.05 and 0.1) in the temperature range of 295 to 90 K are shown in Figure 1(a). The solid symbols are the extracted lattice parameter after analyzing the X-ray diffractograms. It can be seen from Figure 1(a) that the lattice parameters for all the doping (x=0, 0.02, 0.05 and 0.1) vary smoothly with temperature. Figure 1(b) confirms the anomalous lattice parameter contraction for x=0.02 at 90 K, similar to the reported anomaly at room temperature [16]. The solid lines in Figure 1(a) are fit to the equation $a(T) = a(0)\left(1 + \frac{b e^{c/T}}{T(e^{c/T}-1)^2}\right)$, [29] where a(0) is the lattice constant at 0 K, and b and c are fitting parameters.

## 3.2 Raman spectra at ambient condition

The pyrochlore $A_2B_2O_6O'$ can be viewed by forming tetrahedra networks $A_4O'$ and octahedra $BO_6$ with A, B, O and O', respectively, occupying 16c, 16d, 48f and 8b sites. In the pyrochlore structure, there are only two independent variable parameters: (i) the x-coordinate of the O (48f) and (ii) the lattice parameter a. The structure belongs to space group Fd-3m, and the irreducible representations of optical phonons at the zone center are [30] $\Gamma_{Optical} = A_{1g} + E_g + 2T_{1g} + 4T_{2g} + 3A_{2u} + 3E_u + 7T_{1u} + 4T_{2u}$. Amongst these, $A_{1g}$, $E_g$ and $T_{2g}$ are Raman active modes. Because of the inversion symmetry associated with $Eu^{+3}$ and $Ir^{+4}$ sites, heavy cations are stationary in Raman active modes. Six Raman modes at 307 ($T_{2g}^1$), 347 ($E_g$), 382.2 ($T_{2g}^2$), 510.5 ($A_{1g}$), 541 ($T_{2g}^3$) and 678.4 ($T_{2g}^4$) $cm^{-1}$ for undoped $Eu_2Ir_2O_7$ as well as for doped systems are observed (Figure S2 (a)). The seventh mode seen at 703 $cm^{-1}$, marked as P, is possibly a second-order mode. We were not able to distinguish clearly $T_{2g}^4$ and P modes in the doped samples. As mentioned before, the Raman modes in $Eu_2Ir_2O_7$ involve only oxygen atoms in Ir-O, Eu-O and Eu-O' stretching and Ir-O-Ir bending vibrations. The stretching of Eu-O bonds dominates the low-frequency $T_{2g}^1$ and $T_{2g}^2$ modes in $Eu_2Ir_2O_7$. The $T_{2g}^3$ mode involves the motion of Eu-O' stretching, and the high-frequency $T_{2g}^4$ mode involves predominantly Ir-O stretching. The $E_g$ and $A_{1g}$ modes are associated with the Ir-O-Ir bending [26]. According to the study of Bae *et al.* [31] on $Y_2Ru_2O_7$, the $A_{1g}$ mode changes Ru-O-Ru bond angles in-phase (isotropic), whereas changes are not in-phase for the $E_g$ and $T_{2g}$ modes. The doping dependence of the modes is shown in Figure S2 (b). We have also plotted in Figure S2(c) the doping dependence of Ir-O bond length and Ir-O-Ir angle (obtained from the Rietveld analysis of X-ray diffraction data) at ambient temperature to see the correlation between the mode frequencies and structural parameters. The two bending modes, $E_g$ and $A_{1g}$ are influenced differently by doping. It can be seen that there is no simple correlation which suggests that the mode frequencies are also renormalized due to electron-phonon coupling in the metallic state.

## 3.3 Raman studies at low temperatures

Figures 2(a) and (b) show Raman spectra for undoped and doped $(Eu_{(1-x)}Bi_x)_2Ir_2O_7$ for x=0, 0.02, 0.035, 0.05 and 0.1 at two selected temperatures 300 K (inside cryostat) and 4 K in the spectral range of 150 to 800 $cm^{-1}$. We observe changes in the spectra recorded at 4 K compared to spectra at 300 K. The undoped $Eu_2Ir_2O_7$ shows a new mode marked N1 (strong) at 211 $cm^{-1}$(and some weak modes near N1) appearing below the MIT transition temperature of 122 K. The N1 mode is weak for x=0.035 and very broad for x=0.05. However, it is absent for x=0.1. In order to understand the influence of the MIT on the lattice degrees of freedom, we analyze the temperature-dependent behaviour of each doping in detail.

### 3.3.1 Temperature dependence of Raman modes of undoped $Eu_2Ir_2O_7$

First, we address the phonon anomalies in undoped $Eu_2Ir_2O_7$. Figure 3 shows the temperature dependence of frequency and linewidth of three strong modes $T_{2g}^1$, $T_{2g}^2$ and $A_{1g}$. A dashed line drawn at $T_N$=122 K marks the magnetic ordering temperature, clearly showing anomalies and significant changes below $T_N$. To understand the anomalous changes in the frequency and linewidth of the phonons, we first fit the data at temperatures above $T_N$ with the expected changes based on the anharmonic interactions between phonons. In a simplest cubic anharmonic model, a phonon with frequency ωdecays into two phonons of equal frequencies$\omega/2$, giving the temperature dependence of frequency and linewidth (Γ) as [32]:$\omega(T) = \omega(0) + C.G(\omega(0),T)$, and $\Gamma(T) = \Gamma(0) + D.G(\omega(0),T)$, where $G(\omega(0),T) = \left(1 + \frac{2}{exp\left(\frac{\hbar\omega(0)}{2kT}\right) - 1}\right)$, C and D are phonon-phonon interaction parameters, with $C < 0$ and $D > 0$. We have fitted the observed linewidth and frequency with cubic anharmonic model down to 122 K, shown by solid redlines. The fitted graph is extrapolated to 0 K by the dashed red line to bring out the anomalous behaviour below $T_N$. For the $T_{2g}^2$ mode, the anharmonic model with $C<0$ cannot explain the frequency variation above $T_N$, and hence the solid blue line is a straight line fit as a guide to the eye and extrapolated to 0 K by dotted line.

The following observations are noteworthy: (i) the frequency of the $A_{1g}$ mode shows anomalous temperature dependence: frequency softens with decreasing temperature below $T_N$, (ii) the frequencies of $T_{2g}^1$ and $T_{2g}^2$ modes increase sharply below $T_N$, (iii) the linewidths of $T_{2g}^1$ and $T_{2g}^2$ modes below $T_N$ decrease much more than predicted by the anharmonic model, (iv) the frequency of the $T_{2g}^2$ mode increases above $T_N$, anomalous trend opposite to the anharmonic and quasi-harmonic effects seen for the other two modes. We may note that our Raman experiments (unpolarized) on polycrystalline samples do not allow us to isolate the $E_g$ mode from the nearby $T_{2g}^1$ and $T_{2g}^2$ modes. As the $E_g$ mode is shown to red shift with decreasing temperature by ~35 $cm^{-1}$ [26], it gets merged with the $T_{2g}^1$ mode and hence is not seen in 4 K spectrum (bottom most panel of Figure 2(b)). The observations (i) to (iii) point to strong magneto-elastic (spin-phonon) coupling in the AIAO phase. The observation (iv), namely the frequency increase with increasing temperature above $T_N$, can only arise due to strong electron-phonon interaction of the mode in the metallic phase.

### 3.3.2 Raman anomalies for x=0.02

For x=0.02, the $T_N$ decreases to 115 K. Figure4 shows the temperature dependence of frequency and linewidth of the three vibrational modes, which is like the case of undoped $Eu_2Ir_2O_7$. The additional features in the x=0.02 system are: (i) the frequency of the $T_{2g}^1$ mode above $T_N$ is also anomalous, similar to the $T_{2g}^2$ mode, (ii) the linewidth of the $A_{1g}$ mode is anomalous below $T_N$, i.e. it increases as T decreases, and most notably, (iii) the frequencies and linewidths show different temperature dependence below ~50 K, which we identify as T*, a crossover temperature in the AIAO phase [16].

### 3.3.3 Raman anomalies for x=0.035

Figure 5 shows the temperature dependence of phonon frequency and linewidth for the three prominent modes, which brings out the following differences with respect to the temperature dependence of the corresponding modes for x=0 and 0.02: The frequencies of all the three modes show change in their temperature dependence around 170 K. We note that the thermopower changes sign from being positive above 170 K to negative below this temperature [16]. This can change the electron-phonon coupling, and hence the non-monotonic temperature dependence of phonon frequencies above $T_N$ is the combined effect of electron-phonon interaction and anharmonic interactions. The solid lines are fit to the cubic anharmonic model between 300 K and 170 K. Like x=0 and x=0.02, the $A_{1g}$ mode frequency decreases and linewidth increases anomalously below $T_N$.

### 3.3.4 Raman spectra for x=0.05

Figure6 shows the temperature dependence of frequency and linewidth of the three phonons. The frequencies of both low-frequency $T_{2g}$ modes are significantly anomalous above $T_N$ (~63K), shown by solid blue lines, due to the strong electron-phonon coupling in the metallic state. The $A_{1g}$ mode again shows anomalous softening below $T_N$ due to spin-phonon coupling.

### 3.3.5 Temperature dependence of Raman spectra for x=0.1

For x=0.1, the system remains metallic over the entire temperature range and shows Pauli paramagnetism [16]. The cubic anharmonic model in the entire temperature range can capture the trend of the $A_{1g}$ mode, which hardens by ~2 $cm^{-1}$ in frequency and becomes sharper at low temperatures (Figure7). The $T_{2g}$ modes do not show much change with temperature.

## 4 DISCUSSION

Before discussing the temperature dependence of various Raman modes, let us briefly recall the phase diagram of $(Eu_{(1-x)}Bi_x)_2Ir_2O_7$ series previously reported in Ref. [16]. In this series, in the region $0 \leq x \leq 0.035$, the lattice parameter shows an anomalous contraction, and for $x \geq 0.1$, the lattice expands normally. The intermediate region 0.035< x<0.1 is called crossover region. For convenience of the reader, in Figure S3 of the Supplementary information, the magnetic susceptibility and resistivity of the samples are replotted from Ref. [16]. As shown in Figure S3, samples x = 0, 0.02 and 0.035 show a robust transition to a magnetically ordered phase below $T_N$ = 122 K (x = 0), 115 K (x = 0.02), and 108 K (x = 0.035). For these samples, the resistivity also exhibits a sharp MIT at the same respective temperatures. For x = 0.1, on the other hand, no long-range ordering could be seen down to the lowest measured temperature, and it also shows a metallic behavior over the entire temperature range. The behavior of x = 0.05, which lies in the crossover region, is intermediate between these two limits: while the susceptibility of this sample shows a weak Zero-Field-Cooled (ZFC) / Field-cooled (FC) splitting below ≅60 K, the resistivity shows a broad and shallow minimum centered around the same temperature.

We first focus on the softening of the $A_{1g}$ mode below $T_N$ in x = 0, 0.02, and 0.035. To qualitatively understand the frequency softening of the $A_{1g}$ mode vis a vis the frequency hardening of the $T_{2g}$ modes below $T_N$, we consider different contributions to spin-phonon coupling. The spin Hamiltonian can be written as [27], $H = \sum_{i,j}[J_{IE}(S_i . S_j) + D_{ij} . (S_i \times S_j)]$, where $S_i$ and $S_j$ are the $j_{eff}$ = 1/2 spins of $Ir^{4+}$ at i and $j^{th}$ positions in $IrO_6$ octahedra. Here the first term is the Heisenberg IE interaction, and the second term is the DM interaction. The sign of D represents the chirality of spin texture. The effect of the anisotropic exchange interaction and single-ion anisotropy are negligible for $J_{eff}$ = 1/2 related materials [27]. The spin-phonon coupling arises due to the dynamic modulation of the coefficients $J_{IE}$ and $D_{ij}$. The effect of spin-phonon coupling due to the IE interaction has been observed in 3d multiferroic manganates [33], chromates [34–36], and 4d pyrochlore oxides [31]. The contribution of the IE interaction to the renormalization of a given phonon frequency arises from the modulation of $J_{IE}$ by Ir-O vibrational amplitude (u) and Ir-O-Ir bond angle(θ). This is given by $\Delta\omega \sim \lambda_{IE}(S_i . S_j)$, where $\lambda_{IE}$ is proportional to $\frac{\partial^2 J_{IE}(\theta)}{\partial\theta^2}$ and $\frac{\partial^2 J_{IE}(u)}{\partial u^2}$ [27,36]. The sign of $\lambda_{IE}$ can be either positive or negative. In the case of DM interaction, the spin-phonon coupling arises due to modulation of $D_{ij}$ with respect to the Ir-O-Ir bond angle(θ). The phonon renormalization Δω is proportional to $\sum_{i,j} \frac{\partial^2 D_{ij}(\theta)}{\partial\theta^2} . \left(S_i \times S_j\right)$. The relative contributions of IE and DM due to modulation of respective exchange constants have been examined theoretically for $Y_2Ir_2O_7$ [27] and shown that for θ ~130°, the DM contributes an order of magnitude more to the phonon frequency renormalization than the IE.

In the light of the above discussion, let us look at our results shown in Figures 3-5. The anomalous softening below $T_N$ of the $A_{1g}$ mode associated with the Ir-O-Ir bending vibrations is qualitatively similar to the observed softening of another bending vibrational $E_g$ mode in the same system [26], though the magnitude of softening for the $E_g$ mode (~10%) is much higher than that of the $A_{1g}$ mode (~0.5%). This softening arises due to the spin-phonon coupling driven dominantly by the DM interaction with negative sign of $\frac{\partial^2 D_{ij}(\theta)}{\partial\theta^2}$. Similarly, the positive sign of $\lambda_{IE}$ will result in hardening of the $T_{2g}$ stretching modes below $T_N$ for x = 0 and 0.02. The change of behavior below T* ~ 50 K is possibly driven by the WSM ground state. As mentioned before, undoped and doped europium iridates show the WSM phase at low temperatures which will contribute to phonon renormalization [37,38]. In the gapless WSM state, the electron-hole excitation by the phonon with frequency $\omega(T)$ becomes possible resulting in significant electron-phonon interaction. This contributes to additional softening of the phonon frequency and additional linewidth proportional to $H(\omega(0),T) = \left(f(-\frac{\hbar\omega(0)}{2}) - f(\frac{\hbar\omega(0)}{2})\right)$ where f is the Fermi function. The temperature dependence of frequency and linewidth below 50 K for x = 0.02 is fitted (shown by green line Fig. 4) with $\omega(T) = \omega(0) + C.G(\omega(0),T) - A.H\,(\omega(0),T)$, $\Gamma(T) = \Gamma(0) + D.G(\omega(0),T) + B.H(\omega(0),T)$, where A and B are positive constants. At x=0.035 also, the influence of the WSM state below T* can be seen on phonon frequencies and linewidths, similar to the doped system with x = 0.02. The anomalous changes in frequency are highlighted by fitting straight lines shown in blue colour (Fig. 5). In x = 0.035, the hardening of $T_{2g}$ modes is seen only at low temperatures, and in x = 0.05 no hardening could be seen. This relates to the lowering of $T_N$ or weakening of $\lambda_{IE}$ due to increasing bandwidth which is reflected in the increasing electrical conductivity (See Fig. S3).

Next, we discuss the large decrease of the linewidths of the $T_{2g}^1$ and $T_{2g}^2$ modes below $T_N$. This anomalous decrease is most prominent in x = 0 and 0.02. This decrease can be attributed to the absence of electron-phonon coupling in the insulating AIAO phase, which also results in their corresponding frequency increase by a large amount. To sum up this part, our results point out the role of (a) DM interaction in the anomalous temperature dependence of the $A_{1g}$ mode, (b) IE interaction for the stretching $T_{2g}$ modes which changes the bond length and (c) strong electron-phonon coupling in the metallic and the WSM phases.

As shown in Figure 2(b), a new mode marked N1 near 211 $cm^{-1}$ (and some weak modes nearby) is seen below the AIAO transition temperature in x = 0, 0.02 and 0.035, i.e., the samples where a sharp magnetic transition to AIAO phase is observed. In x = 0.05, where the magnetic ground state has weakened considerably, the mode N1 is indistinguishable from the background; and finally, in x = 0.1, that does not show any magnetic ordering, no signs of N1 could be seen. Clearly, presence of N1 can be associated with a robust magnetically ordered phase. By comparing inelastic X-ray scattering experiments, it has been suggested that the low-frequency mode N1 is coming from single magnon excitation [26,39]. The temperature dependence of the N1 mode frequency and its intensity relative to the $T_{2g}^2$ phonon are plotted in Figure S4 (a) and S4 (b) for x = 0 and 0.02, respectively. A slight hardening of N1 with x can be attributed to the higher spin-exchange constant expected due to lattice contraction. Taking the intensity of N1 mode as an order parameter, the red line is a fit to ~ $(T_N\text{-}T)^{½}$, as expected in the mean-field theory.

## 5 CONCLUSIONS

Our Raman studies have shown that the Ir-O-Ir bending mode $A_{1g}$ and $T_{2g}^1$ and $T_{2g}^2$ modes, which have some contributions from the Ir-O stretching vibrations, are significantly influenced by strong spin-phonon coupling below $T_N$ and strong electron-phonon coupling above $T_N$. Further, the emergence of the Weyl semi-metal phase (WSM) at low temperatures (below T*~50 K) contributes to a noticeable phonon renormalization. These results, not reported so far for the Eu-iridate systems, add to the exploration of novel properties of pyrochlore iridates arising from spin-phonon and electron-phonon interactions. We hope that our detailed experimental results will motivate further theoretical calculations to understand phonon anomalies quantitatively in the AIAO phase as well as above $T_N$.

## ACKNOWLEDGEMENTS

AKS thanks the Department of Science and Technology, Government of India, for financial support under the Year of Science Professorship and Nano Mission Council. AT acknowledges support from the Council for Scientific and Industrial Research (CSIR), India. S.S. would like to thank the Department of Science and Technology (DST), India and Science and Engineering Research Board (SERB), India for financial support under Grant No. EMR/2016/003792. P.T.

and S.S. acknowledge DST for travel grant for performing low-temperature synchrotron x-ray diffraction. P.T. and S.S. acknowledge DESY (Hamburg, Germany), a member of the Helmholtz Association HGF, for the provision of X-Ray diffraction experimental facilities at Beamline P02 at PETRA III.

**LIST OF FIGURES**

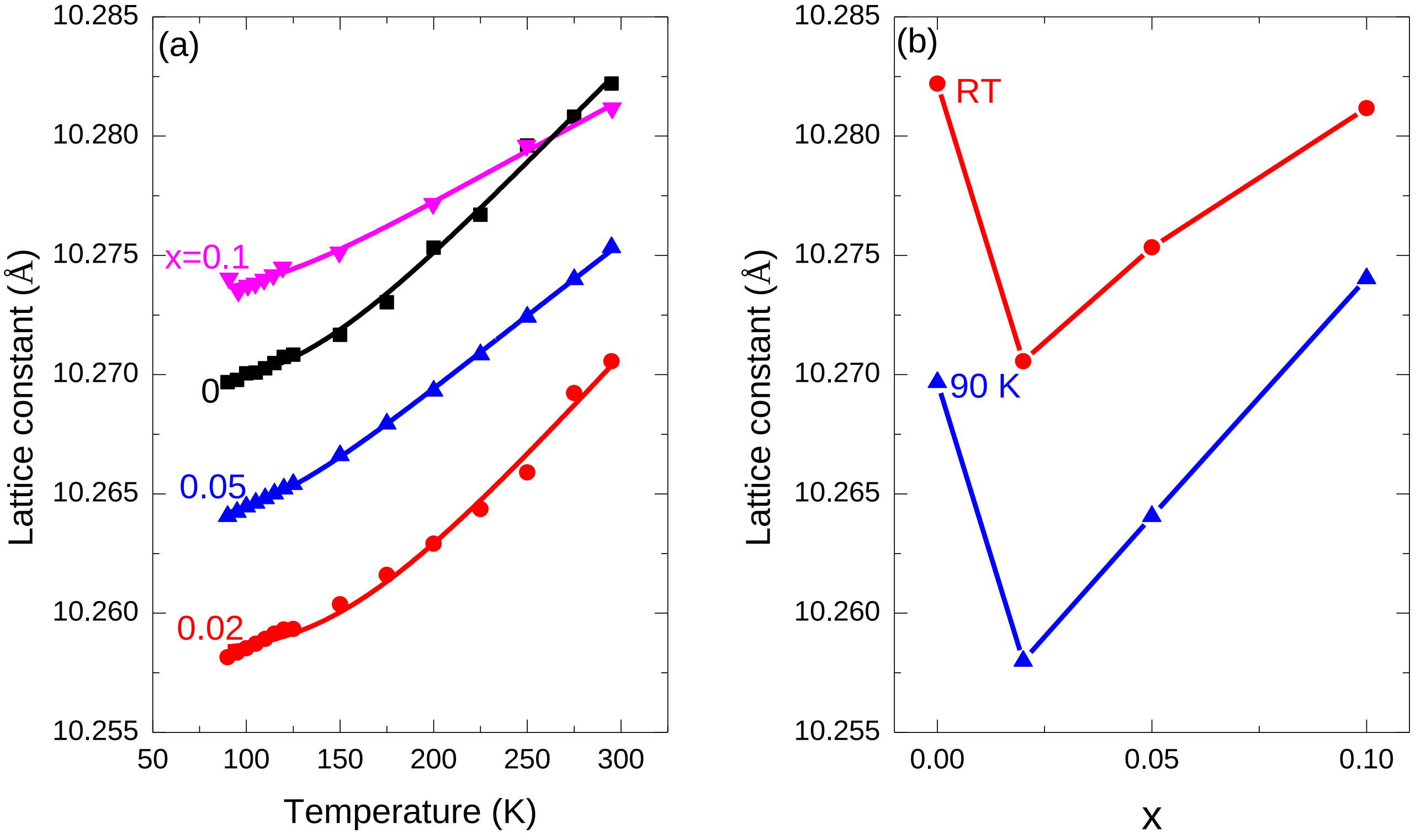

FIG.1. (a) Variation of the lattice parameter of $(Eu_{(1-x)}Bi_x)_2Ir_2O_7$ with temperature. Symbols represent lattice parameter. Solid lines are fit to the data as discussed in the text. (b) Lattice constant vs x at room temperature (RT) and 90 K.

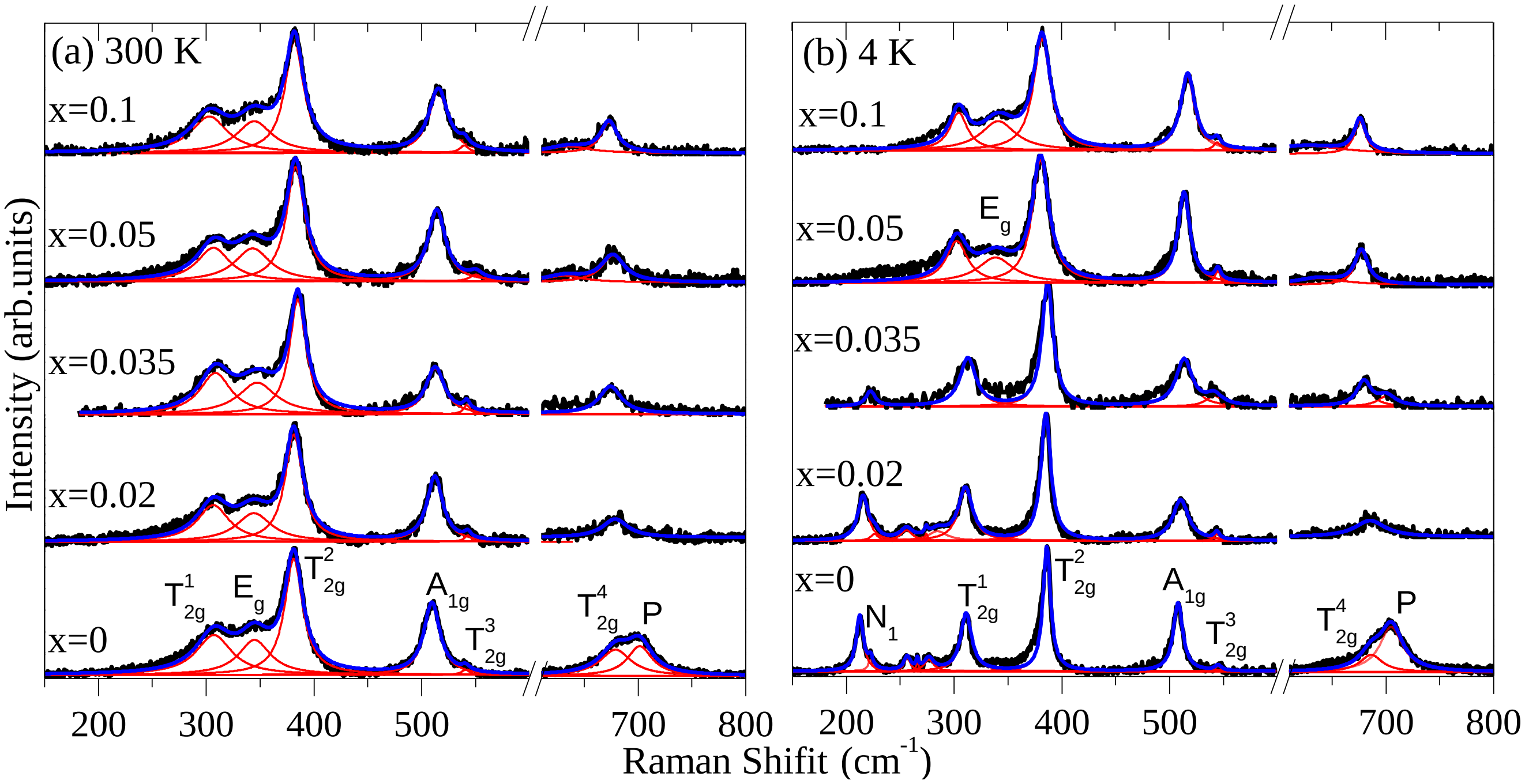

FIG.2. Raman Spectra of $(Eu_{(1-x)}Bi_x)_2Ir_2O_7$ at (a) 300K and (b) 4 K for x = 0, 0.02, 0.035, 0.05 and 0.1. The thick black line represents raw data that are fitted with a sum of Lorentzian functions (shown by blue lines). The thin red line represents individually fitted phonon modes.

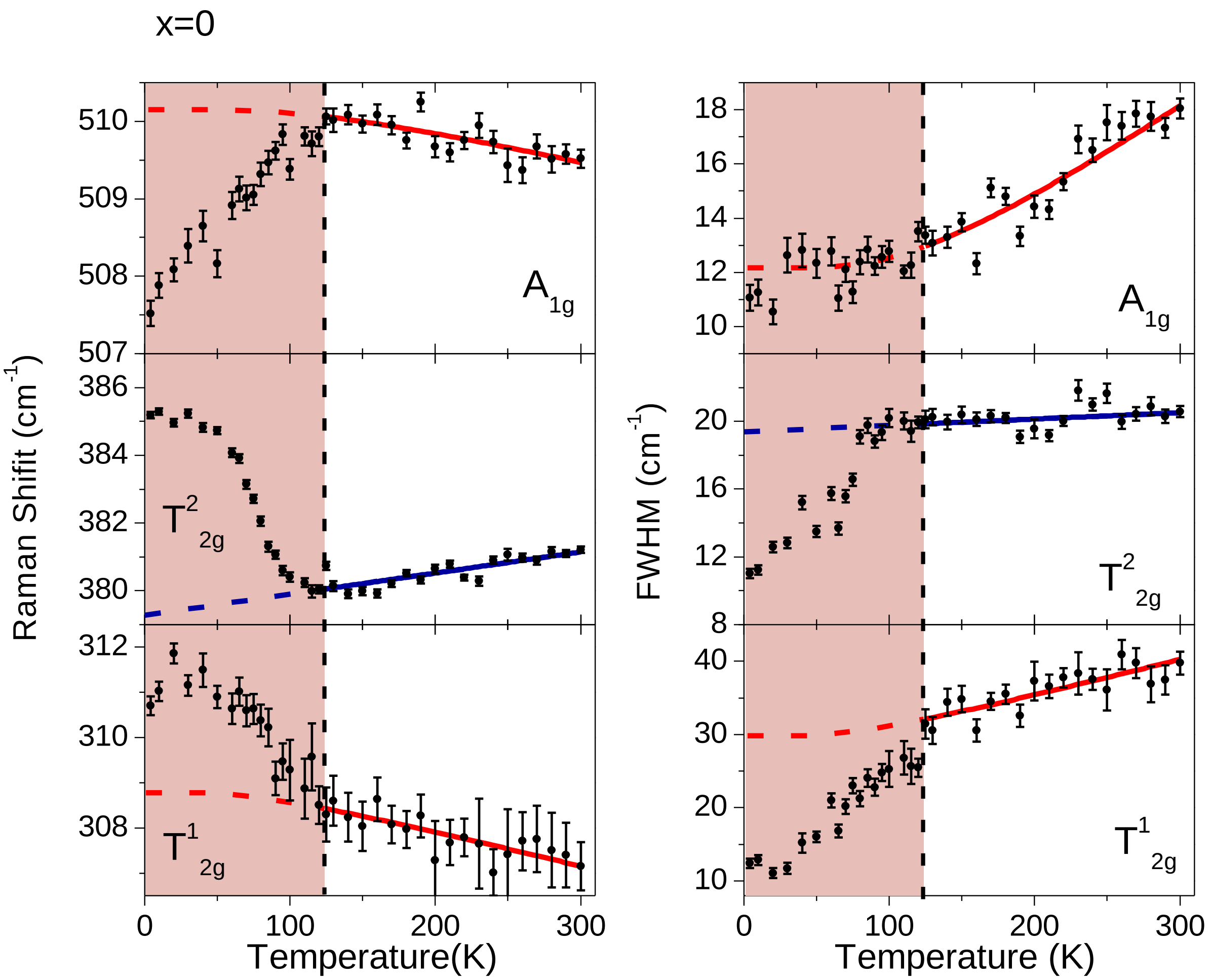

FIG.3. Temperature dependence of frequency and linewidth of Raman modes $T_{2g}^1$, $T_{2g}^2$, and $A_{1g}$ for $Eu_2Ir_2O_7$. A vertical dashed line is drawn at $T_N$ =122 K to mark the metal-insulator transition temperature. Solid red lines are the fits of the cubic anharmonic model to the data above $T_N$, extrapolated to 0 K by the dotted lines. Blue lines are linear fits to the data as a guide to the eye.

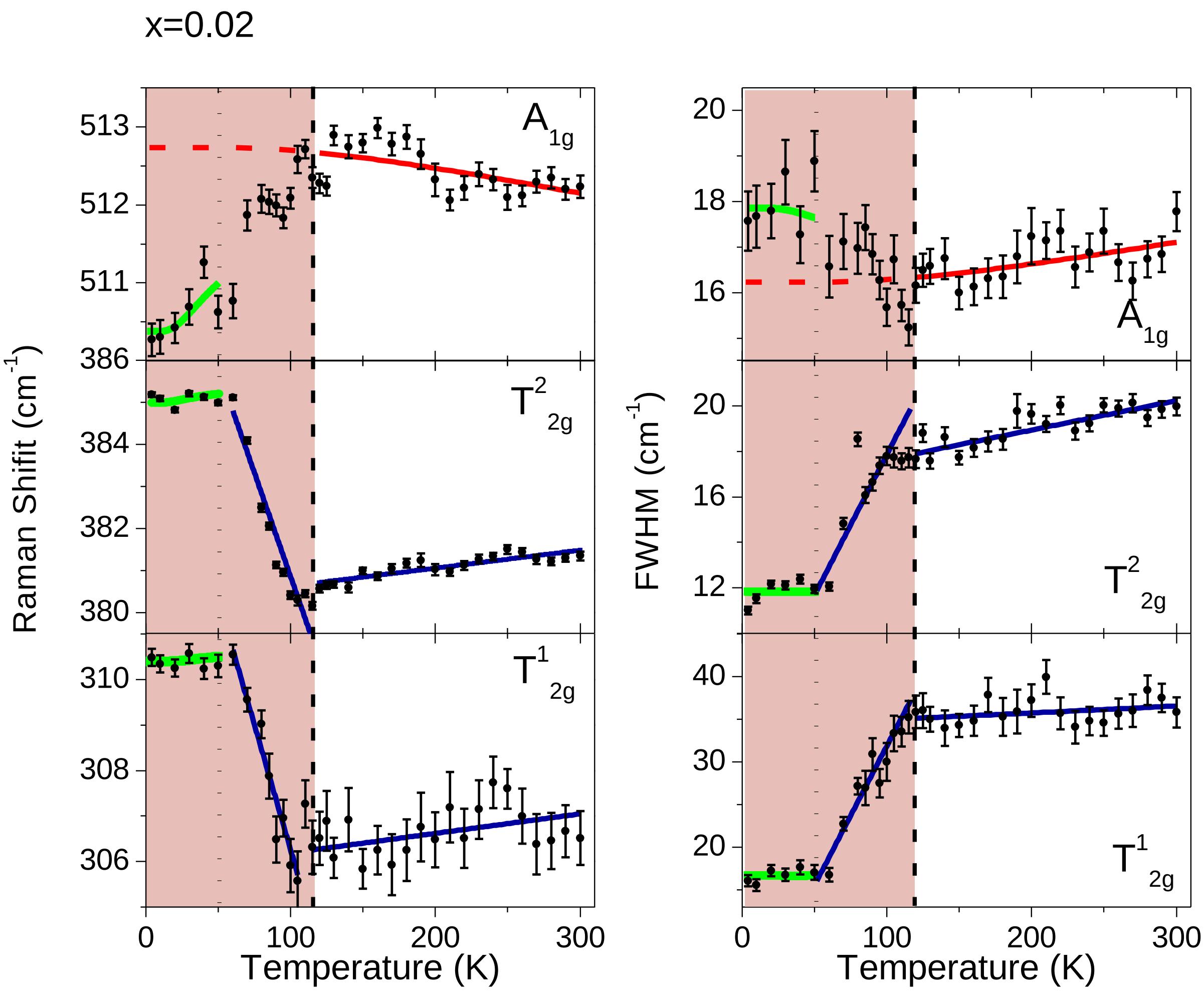

FIG.4. Temperature dependence of frequency and linewidth of Raman modes $T_{2g}^{1}$, $T_{2g}^{2}$, and $A_{1g}$ for x=0.02. A vertical dashed line marks $T_N$ =115 K. A vertical dotted line at 50 K marks T*. The solid red and blue lines are similar to Figure 3. The green lines below T* take into account electron-phonon contributions to the phonon renormalization in the WSM state.

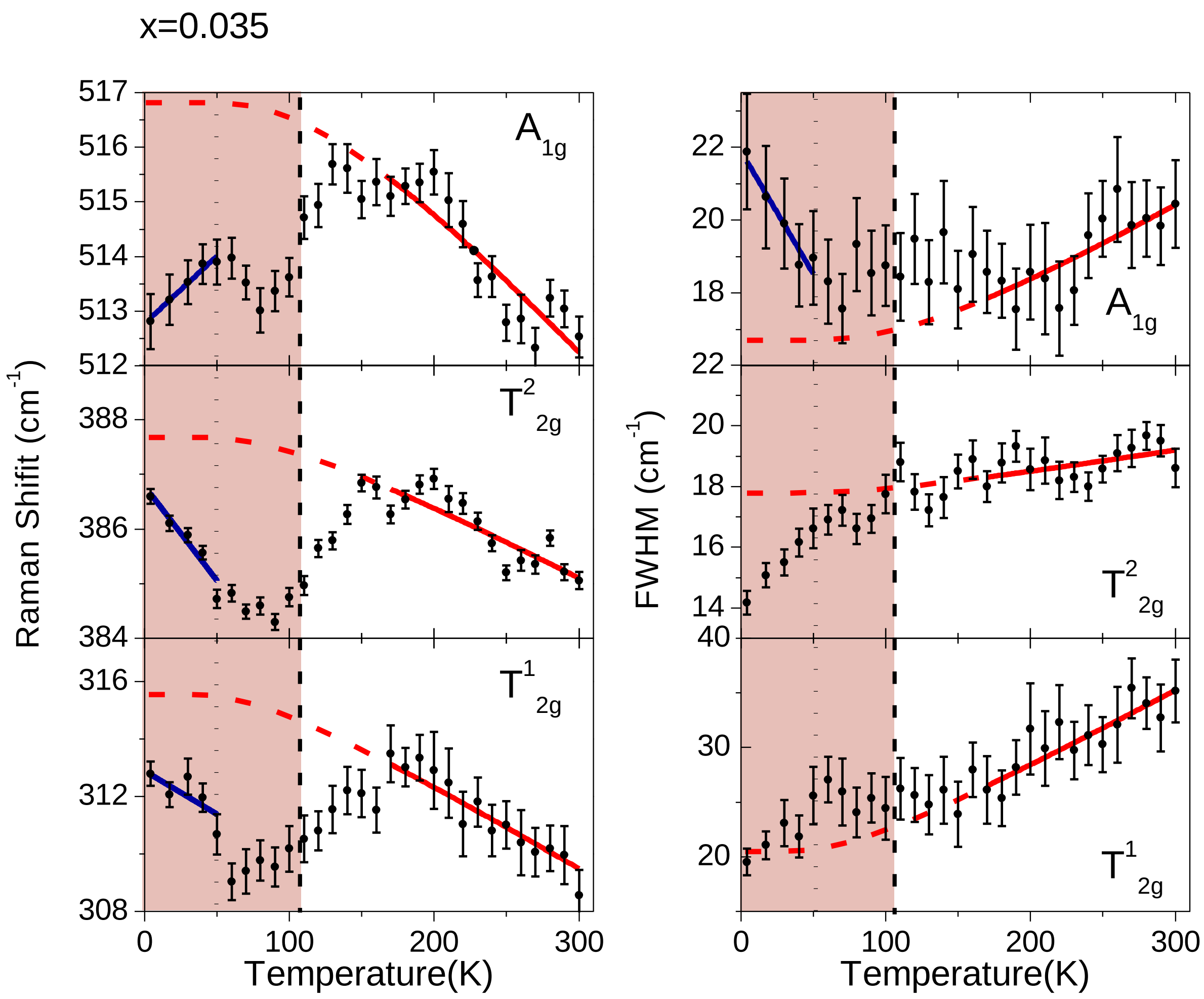

FIG.5. Temperature dependence of frequency and linewidth of Raman modes $T_{2g}^{1}$, $T_{2g}^{2}$, and $A_{1g}$ for x=0.035. A vertical dashed line is at $T_N$ =108 K. A vertical dotted line at 50 K marks T*. The meaning of lines is the same as in Figure 3.

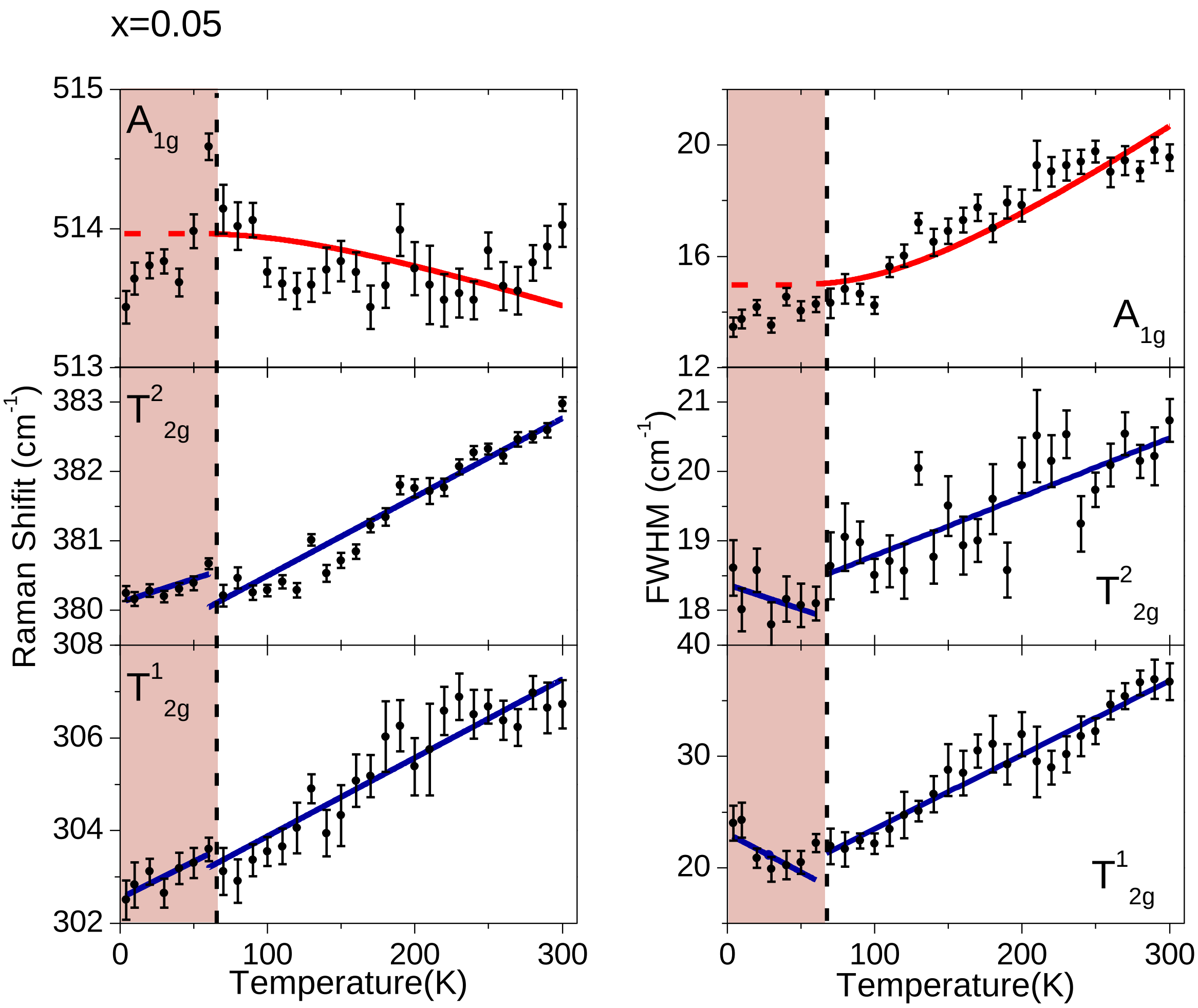

FIG.6. Temperature dependence of frequency and linewidth of Raman modes $T_{2g}{}^1$, $T_{2g}{}^2$, and $A_{1g}$ for x=0.05. A vertical dashed line marks $T_N$ =63 K. The meaning of solid lines are the same as in Figure 3.

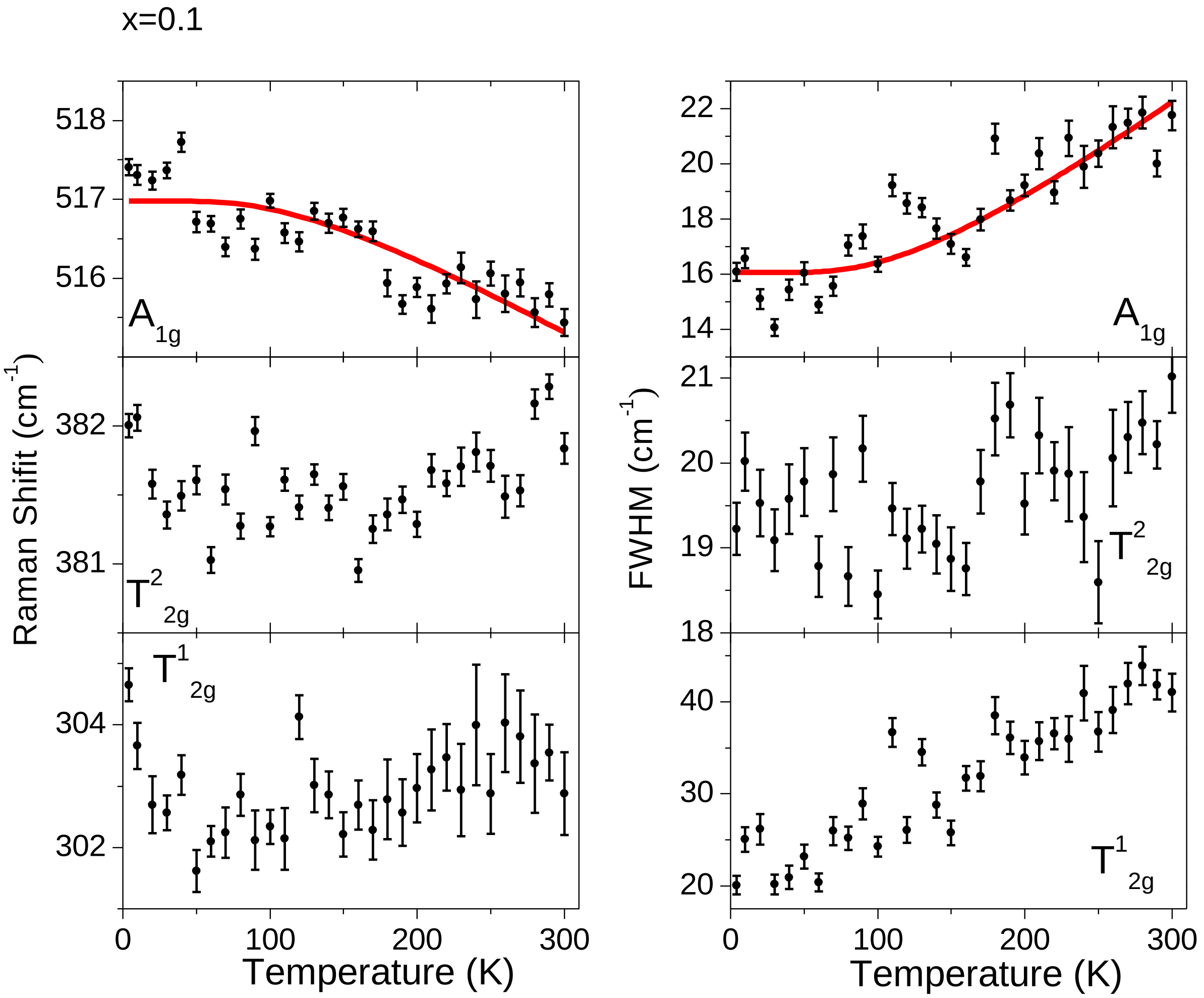

FIG.7. Temperature dependence of frequency and linewidth of Raman modes $T_{2g}^{1}$, $T_{2g}^{2}$, and $A_{1g}$ for x = 0.1. The thick red lines are the same as in Figure 3.

**SUPPORTING INFORMATION**

# Role of spin-phonon and electron-phonon interactions in phonon renormalization of $(Eu_{(1-x)}Bi_x)_2Ir_2O_7$ across the metal-insulator phase transition: Temperature-dependent Raman and X-ray studies

Anoop Thomas [1], Prachi Telang [2], Kshiti Mishra [2], Martin Cesnek [3], Jozef Bednarcik [4 #], D V S Muthu [1], Surjeet Singh [2] and A K Sood [1*]

[1] *Department of Physics, Indian Institute of Science, Bangalore, 560012, India*

[2] *Department of Physics, Indian Institute of Science Education and Research, Dr Homi Bhabha Road, Pune 411008, India*

[3] *Department of Nuclear Reactors, Faculty of Nuclear Sciences and Physical Engineering, Czech Technical University in Prague, V Holesovickach 2, 180 00 Prague, Czech Republic*

[4] *DeutschesElektronen-Synchrotron DESY, Notkestrasse 85, D-22607 Hamburg, Germany*

*****Corresponding Author**,** E-mail: asood@iisc.ac.in

[#] Current affiliation - Institute of Physics, Faculty of Science, P.J. Safarik University in Kosice, Park Angelinum 9, 041 54 Kosice, Slovak Republic

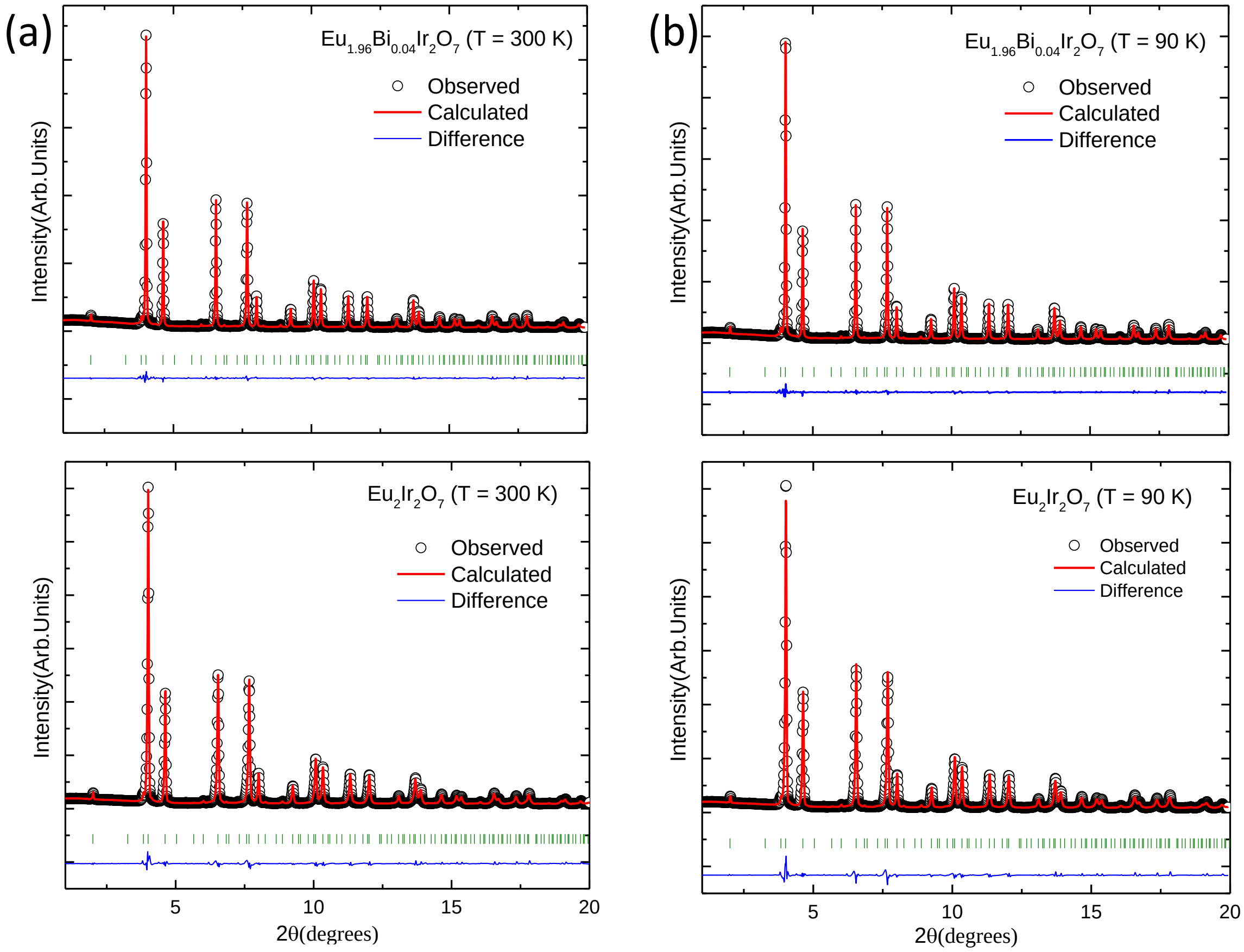

FIG. S1. Representative Rietveld refinement plot for $(Eu_{(1-x)}Bi_x)_2Ir_2O_7$ at (a) 300 K and (b) 90K.

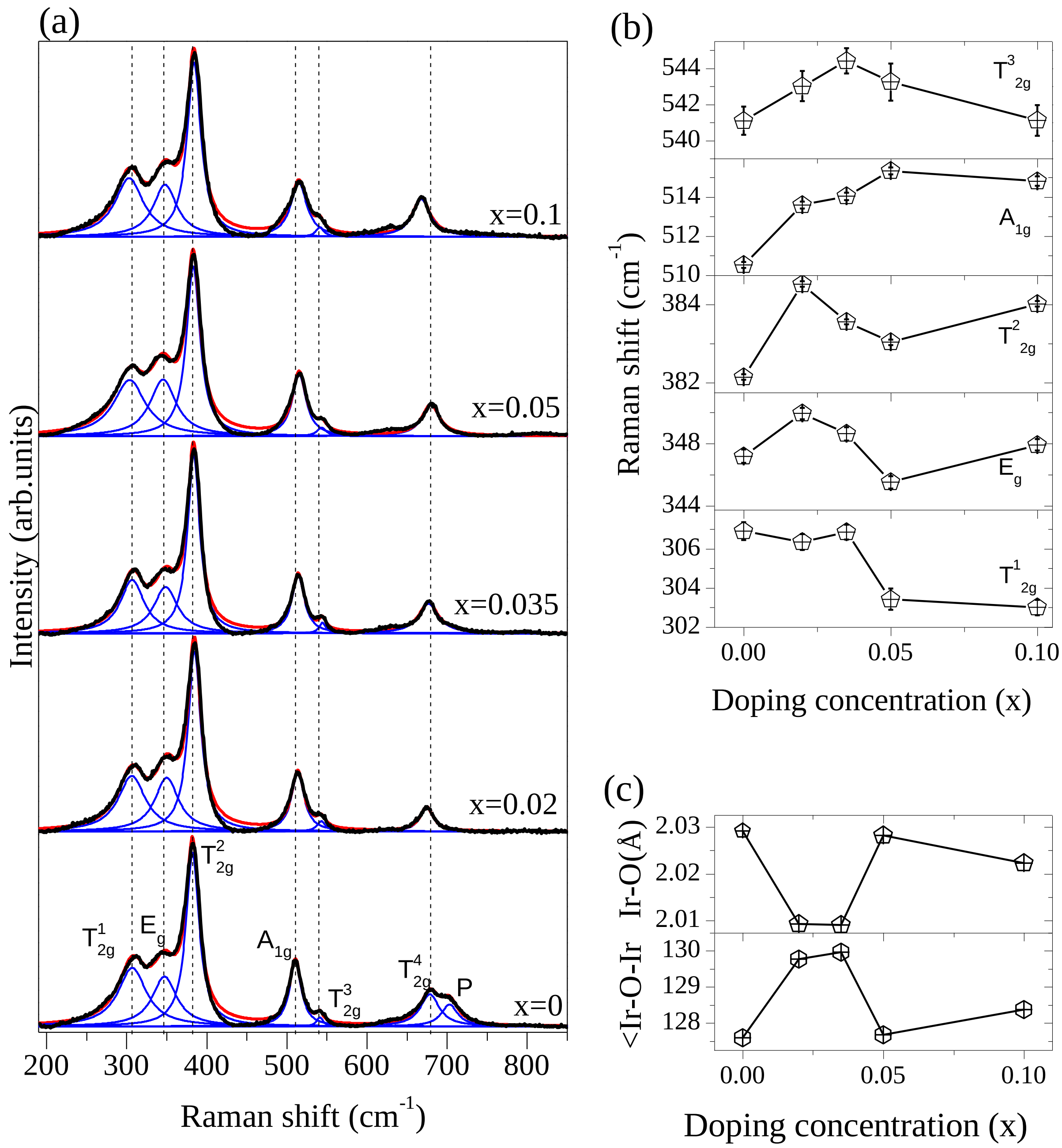

FIG. S2. (a) Raman spectra of $(Eu_{(1-x)}Bi_x)_2Ir_2O_7$ at room temperature. The black line represents raw data that are fitted with a sum of Lorentzian functions (shown by red lines). The blue line represents individually fitted phonon modes. (b) Raman frequencies as a function of doping concentration x. (c) Evolution of Ir-O bond length and Ir-O-Ir bond angle vs x. Solid lines in the panels b and c are a guide to the eye.

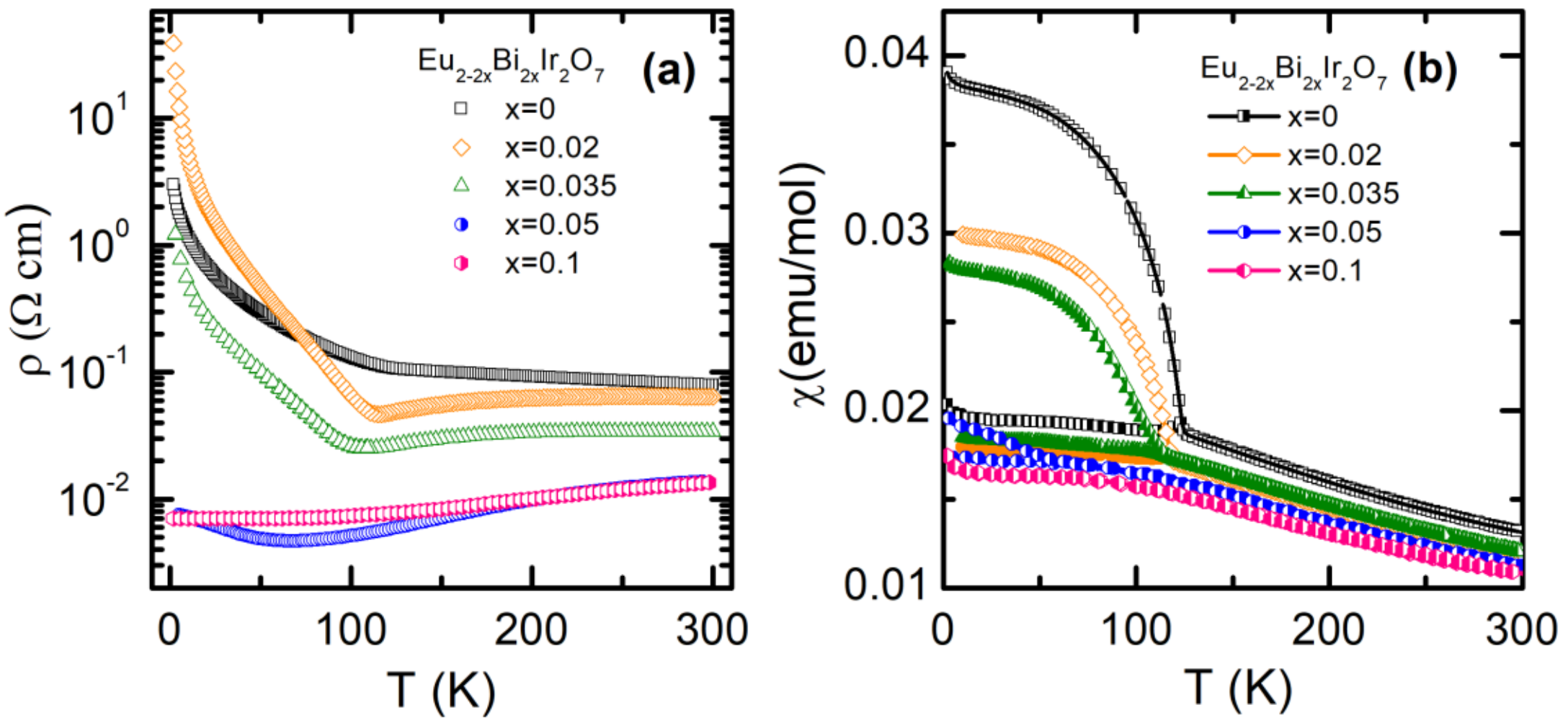

FIG. S3. Temperature variation of (a) electrical resistivity (ρ) and (b) magnetic susceptibility (χ) of the samples investigated here (adapted from Ref. [16])

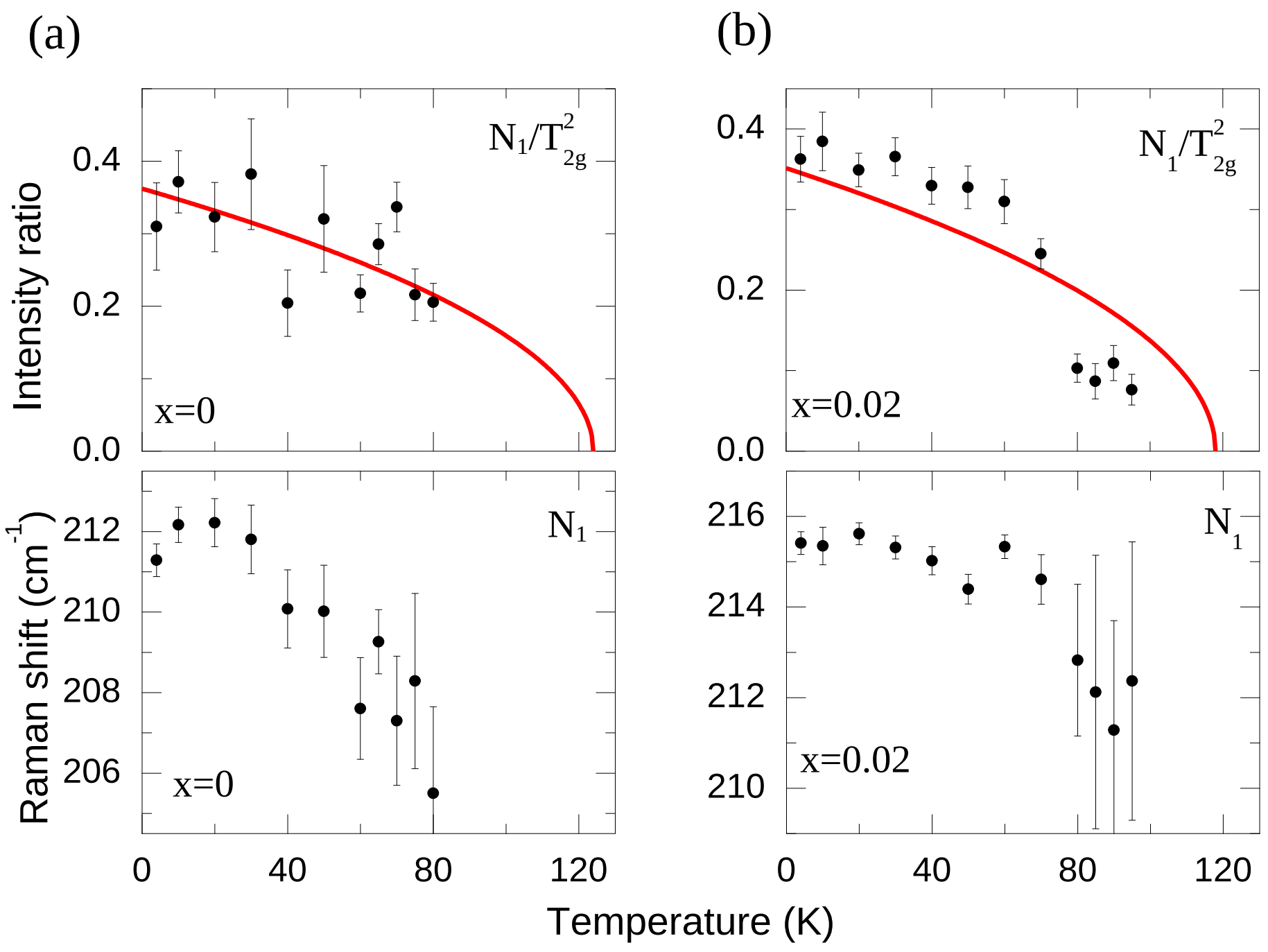

FIG. S4. Temperature dependence of frequency (bottom panels) and relative intensities (top panels) of the one magnon Raman mode N1 for (a) x = 0 and (b) x = 0.02. The red solid lines are fit to the experimental data as discussed in the text.